\documentclass[DIV14,a4paper]{scrartcl}

\usepackage[english]{babel}
\usepackage[pdftex]{graphicx}
\usepackage[T1]{fontenc}
\usepackage[ansinew]{inputenc}
\usepackage{lmodern}

\usepackage{mathptmx}

\usepackage{color}
\usepackage[caption=false]{subfig}
\usepackage[hyphens]{url}
\usepackage{cite}
\usepackage{rotating} 
\usepackage{fancyhdr}

\usepackage{listings}
\newcommand{\todobi}{\begin{itemize} \itemsep0em \footnotesize}
\newcommand{\todoe}{\end{itemize}}

\newcommand{\bi}{\begin{itemize}}
\newcommand{\ei}{\end{itemize}}

\newcommand{\mc}[1]{\multicolumn{2}{c}{\textbf{#1}}}

\newcommand{\fn}[1]{\textsuperscript{#1}}

\usepackage[
  a4paper,
  colorlinks=true,
  linkcolor=blue,
  citecolor=blue,
  urlcolor=blue,
  bookmarksopen=true,
  bookmarksnumbered=true
]{hyperref}

\begin{document}

\title{Short Note on Costs of Floating Point Operations on current x86-64 Architectures:
Denormals, Overflow, Underflow, and Division by Zero}

\author{Markus Wittmann \and
Thomas Zeiser \and
Georg Hager \and
Gerhard Wellein}
\publishers{\small
  Erlangen Regional Computing Center \\
  University of Erlangen-N\"urnberg \\
  Martensstra{\ss}e 1, 91058 Erlangen, Germany \\
  hpc@fau.de
}
\date{}

\maketitle  

\abstract{
Simple floating point operations like addition or multiplication on normalized
floating point values can be computed by current AMD and Intel
processors in three to five cycles. 
This is different for denormalized numbers,
which appear when an underflow occurs and the value can no longer be represented as a
normalized floating-point value. 
Here the costs are about two magnitudes larger.
}

\section{Introduction}

Simple floating point operations like addition or multiplication on normalized
floating point values can nowadays be computed by current AMD and Intel
processors in three to five cycles. 
This is different for denormalized numbers,
which appear when an underflow occurs and the value can no longer be represented as a
normalized floating-point value. 
Here the costs are about two magnitudes larger.
Often this is not noticed as this \textit{gradual underflow} is normally
avoided, by configuring the floating point units to tread underflowed values as
zero, as described in section~\ref{sec:ftz-daz}.

The object of this short report is to quantify the performance impact on
floating point operations when denormalized/NaN values, overflows, or
divisions by zero occur.
Hereby the focus is only on 
\bi
  \item double precision floating point addition, multiplication, division and
        fused-multiply-add
  \item with the AVX, AVX2, and FMA3/4 ISA extensions
\ei
for the x86-64 architecture. 
Single precision, x87, SSE, the influence of different rounding modes, etc. are 
not considered.

\section{Flush-to-Zero and Denormals-are-Zero}
\label{sec:ftz-daz}

The SSE/AVX floating point units of the current x86-64 architecture support two
complimentary modes for avoiding the enormous costs of gradual underflow:
\bi
  \item DAZ: Denormalized values of input operants are treated as
        zero, which is called \textit{denormals are zero}.
  \item FTZ: With \textit{flush to zero} (FTZ) the
        denormalized result of floating point operations are set to zero.
\ei
Both options are controlled through specific bits in the floating
point control register \texttt{MXCSR}.
For FTZ and DAZ bit $15$ and $6$ is responsible, respectively.
Additionally for FTZ it makes sense to mask underflow exceptions through bit
$11$.
The manipulation of \texttt{MXCSR} is performed via the \texttt{LDMXCSR} and
\texttt{STMXCSR} instructions or their intrinsic equivalents
\texttt{\_mm\_getcsr()} and \texttt{\_mm\_setcsr()}.
Intel provides some more details~\cite{intel-ftz-daz}.
Per default GCC and Intel compiler insert code to use FTZ and DAZ, which can be
altered via parameters.
This is described in the corresponding compiler documentation.

\section{Benchmarked Systems}

For measuring the duration of floating point operations three Intel-based
and one AMD-based system were used.
The Intel systems are based on the three microarchitectures SandyBridge, IvyBridge,
and Haswell.
From AMD only the older Bullzoder-based Interlagos was available.
Table~\ref{tab:cpus} gives a short overview of the systems' parameter.
Instruction throughput and latency numbers are taken from the
vendors~\cite{intel-orm-2014,amd-guide-2014} and \textsc{Fog}~\cite{fog-2014}. 
Throughput describes how many independent instructions of a certain type
can be issued per cycle.
On the other hand latency denotes the duration of the execution of an
instruction in cycles.

On all Intel-based processors each core has a separate multiplication and
addition unit.
This enables them to execute these two operations in parallel.
Each Haswell core has an additional multiplication unit, located on the same port
as the add unit. 
Either two multiplications or one addition and multiplication can be
executed concurrently.
Additionally, two $256$-bit wide FMA units are available~\cite{intel-orm-2014},
sharing the same ports as the multiplication and addition ports.

An Interlagos floating point module instead, which is shared by two adjacent
cores, has two $128$-bit wide fused-multiply-add (FMA)
units~\cite{amd-guide-2014}.
On each cycle they can receive an AVX multiplication, addition, division, or FMA
from one of both cores.

The FMA support from Intel and AMD differs in their implementation.
Intel uses a three operant destructive form (FMA3): $a = a \times b + c$.
AMD on the other hand uses four operants (FMA4) where the source operant
is not overridden: $a = b \times c + d$.

\begin{table}
  \begin{center}
  \begin{minipage}{\textwidth}\centering
    \begin{tabular}{lll|rrrrrrrr}
      \hline
      &
      && \multicolumn{1}{c}{\textbf{SandyBridge}} 
      && \multicolumn{1}{c}{\textbf{IvyBridge}} 
      && \multicolumn{1}{c}{\textbf{Haswell}}   
      && \multicolumn{1}{c}{\textbf{Interlagos}} \\
      \hline
      Type              &         && Intel          && Intel      && Intel        && AMD \\
                        &         &&  Xeon E5-2680  && E5-2660 v2 && E5-2695 v3   && Opteron 6276 \\

      Frequency  & [GHz] && 2.7  && 2.2  && 2.3 &&  2.3 \\
      Cores      &       && 8    && 10   &&  12 &&   16   \\
      ISA        &       && AVX  && AVX  && AVX, AVX2 && AVX, FMA4 \\  
                 &       &&      &&      &&      FMA3 &&  \\  
      \hline
      AVX Addition       & per cy && 1 && 1  && 1 && 1 \\
      AVX Multiplication & per cy && 1 && 1  && 2 && 1 \\
      AVX Add/Mul        & per cy && 1/1 && 1/1 && 1/1, 0/2&& 1/0, 0/1 \\
      \hline
      AVX Addition &&& \\
      ~~Throughput & per cy &&  1 && 1 && 1 && 1 \\
      ~~Latency    & [cy]   &&  3 && 3 && 3 && 6 \\
      AVX Multiplication   &&& \\
      ~~Throughput & per cy &&  1 && 1 && 2 && 1 \\
      ~~Latency    & [cy]   &&  5 && 5 && 5 && 6 \\
      AVX Division &&& \\
      ~~Throughput & per cy &&  \fn{a}0.025 &&  \fn{a}0.04 &&  \fn{a}0.04 && \fn{b}0.03--0.11 \\
      ~~Latency    & [cy]   &&  \fn{b}21--45 && \fn{b}20--35 && \fn{b}19--35 && 27 \\
      FMA ($256$-bit wide)  &&& \\
      ~~Throughput & per cy && -- && -- && 2 && 1 \\
      ~~Latency    & [cy]   && -- && -- && 5 && 6 \\

      \hline
    \end{tabular}
    \caption{Relevant architectural characteristics of the evaluated systems. If
  not otherwise noted, instruction throughput and latency numbers are taken from
  \cite{intel-orm-2014,amd-guide-2014,fog-2014}.}
    \label{tab:cpus}
  \end{minipage}
\footnoterule\footnotesize
  \begin{minipage}[t]{\textwidth}
  \textsuperscript{a}
        Measured with micro-benchmark.\par 
  \textsuperscript{b}
        Taken from \textsc{Fog}~\cite{fog-2014}.\par
  \end{minipage}\hfill
  \end{center}
\end{table}

\section{Micro-Benchmarks}

For benchmarking small micro-benchmarks were used which perform following
operations on double precision floating point vectors:
%

\begin{tabular}{lll}
\\
 \textbullet ~ Addition:       & & \texttt{a(:) = b(:) + c(:)} \\[0.1cm]
 \textbullet ~ Multiplication: & & \texttt{a(:) = b(:) $\times$ c(:)} \\[0.1cm]
 \textbullet ~ Division:       & & \texttt{a(:) = b(:) / c(:)} \\[0.1cm]
 \textbullet ~ FMA:            & & \texttt{a(:) = b(:) $\times$ c(:) + d(:)}
\\[1ex]
\end{tabular}

Different types of input values are tested.
Firstly the vectors are initialized in such a way that the results of the 
computations are normalized values.
Further input values are chosen, which provoke underflow, overflow, or division
by zero.
And finally it is tested how the duration of the operation is influenced if
not-a-number (NaN) values are used as input operants.
The operations are implemented as two nested loops over the vectors to ensure
the duration of the benchmark is long enough:
\\

\begin{lstlisting}
  for (int n = 0; n < repetitions; ++n) {
    for (int i = 0; i < vectorLength; ++i) {
      // benchmark operation on vector element i
    }
  }
\end{lstlisting}
With AVX the innermost loop gets vectorized, so that during one AVX iteration
four scalar iterations are performed at the same time.

Implementing these operations with C/C++ or Fortran requires beside executing
the computations itself loading and storing the involved vectors.
This introduces a bottleneck, even when the vectors reside in the cores' L1
cache and the full floating point performance will not be visible.
With the shown vector operations all iterations over the vectors are independent
and prefetching, as well as the out-of-order engine, can work perfectly.
Thus for computing the resulting performance only the throughput of the
instructions is relevant and latencies can be ignored, assuming data resides in
the L1 cache.
The SandyBridge and IvyBridge systems have the following properties:
\bi
  \item $1$\,cy for full AVX load,
  \item $2$\,cy for full AVX store,
  \item $1$\,cy for AVX multiplication,
  \item $1$\,cy for AVX addition.
\ei
To perform one AVX iteration, i.\,e.\ four scalar iterations, the multiplication 
and add benchmarks each require
\bi
  \item two AVX loads, 
  \item one AVX store, 
  \item one multiplication/addition.
\ei
All evaluated processors are superscalar and can execute load, store, and
arithmetic instructions concurrently.
With this assumptions one AVX iteration takes $2$\,cy, as it is limited by the 
single AVX store and the two AVX loads.
On the average one single addition or multiplication of the corresponding AVX
version takes $0.5$\,cy.
This can be seen in Tab.~\ref{tab:results} for the corresponding architectures
when the C kernel is used.

The addition and multiplication units however, have a throughput of $1$~AVX
addition/multiplication per cycle, which results in $0.25$\,cy per single
operation.
This limit can only be reached if the bottleneck is removed and the code is no
more load and store bound.

By explicitly implementing these benchmarks in assembly this problem can be
avoided.
The vector length is chosen short enough so that all operants can be kept in
registers, so that additional loads or stores from and to the cache are no longer 
required.
With these benchmarks the full throughput of $0.25$\,cy 
is achieved as reported in Tab.~\ref{tab:results} for normalized numbers with
the ASM kernel. 

The body of the innermost iteration loop of the multiplication benchmark for
example looks like the following with this adjustments (Intel semantics):
\begin{lstlisting}
    vmulpd   ymm9, ymm1, ymm5
    vmulpd  ymm10, ymm2, ymm6
    vmulpd  ymm11, ymm3, ymm7
    vmulpd  ymm12, ymm4, ymm8
\end{lstlisting}
Here a vector length of $16$ was chosen.
The registers $YMM1$--$YMM8$ are initialized with the values of the vectors $a$
and $b$ before the loop is entered and are then reused.
It is important to note that the innermost loop must now be unrolled often
enough to hide the latency of the benchmarked operations.
For the multiplication this is $5$\,cy, as there exists no dependency between 
the target registers.
A unroll factor $\ge 4$ hides this latency as shown in the previous code
snippet.

For the benchmark it is assumed that the execution units are not able to cache
previously computed values over a cycle of the innermost loop and thus cannot
use some short cut when same operants appear again.

All other benchmarks are implemented accordingly using the AVX instruction set.
Additionally for the FMA benchmark on Haswell and Interlagos FMA3 and FMA4 were
used, respectively.

\section{Results}

All micro benchmarks were executed with enabled and disabled FTZ and DAZ.
The results are shown in Tab.~\ref{tab:results}.
The reported values specify the duration of a single floating-point operation in
cycles for the specific operation like addition, multiplication, division, or 
fused-multiply-add.
The visible duration of the full AVX or FMA instruction is four times the
reported number. 
The input values for the micro benchmarks were adjusted to generate as a result 
normalized values, underflows, overflows, or divisions-by-zero.
Furthermore the impact of denormalized and Nan values as input operants are
evaluated.
As already mentioned the full throughput for addition, multiplication,
division-by-zero, and FMA is only reached when utilizing the assembly version of the 
benchmark (ASM) instead of the C implementation.

With FTZ and DAZ enabled (``F+D'' columns) the measured durations of the
specific operations are within the documented ranges.
Disabling FTZ and DAZ (``No F+D'' columns) as expected does not increase the
costs for operations with normalized input and output values.

\paragraph{Addition}
Without FTZ and DAZ the duration of the additions is independent of overflows, 
denormalized input values, and NaNs.
Additions are only sensitive to underflows, which take then around
$36$--$38$\,cy.
Despite these high values, throughout the development of Intel's
microarchitectures it is evident, that the handling of this case has been
improved.
The duration of an underflowing addition was reduced from
$38.20$\,cy (SandyBridge), over $37.70$\,cy (IvyBridge) to $31.90$\,cy
(Haswell).

\paragraph{Multiplication}
Multiplications become expensive in case FTZ and DAZ are disabled
and either an underflow occurs or input operants contain denormalized values.
On the Intel processors, if both input operants are denormals, i.\,e.\ the
multiplicand and the multiplier, then the duration is $0.25$\,cy, the same as
with normalized values.
As in the case of the addition, an overflow or NaN input values introduce no
extra cost.

\paragraph{Division}
The duration of a division ranges from $7$ to $10$\,cy on the Intel
architectures and requires $5$\,cy on Interlagos for normalized input values.
With enabled FTZ and DAZ overflows and underflows to not introduce additional
costs.
Division-by-zero and denormalized input values seem to be detected in an early
stage.
Their throughput duration is only half of a division with normalized operants.

With disabled FTZ and DAZ overflows have no impact on the instruction duration.
In contrast, an underflow in the division takes $71$\,cy (SNB), $63$\,cy (IVB),
and $57$\,cy (HSW) compared to the $\approx 41$\,cy on the AMD system.
Denormalized input operants are always connected with a penalty, except for the
AMD system, where only a denormalized dividend is expensive.

\paragraph{FMA}
According to IEEE 754-2008\cite{ieee-754-2008} the fused-multiply-add operation
should compute $b \times c + d$ as with infinite precision and round only once
at the end.
Haswell with FMA3 and Interlaogs with FMA4 show both an interesting behavior,
when an underflow in the multiplication of the FMA occurs and FTZ and DAZ are 
disabled.
An underflow with a pure AVX multiplication instruction (FTZ and DAZ is
disabled) costs $33$\,cy (Haswell) and $37$\,cy (Interlaogs), whereas no penalty 
is measured, when this occurs with the FMA instructions.
In contrast, an underflowing addition in FMA with disabled FTZ and DAZ is
time-consuming.

\begin{sidewaystable}       
  \footnotesize
  \begin{center}               

  \begin{tabular}{ll|rrrrrrrrrrrrr}
    \hline
                    && \mc{SandyBridge} && \mc{IvyBridge} && \mc{Haswell} && \mc{Interlagos} && \textbf{Kernel} \\
                    &&  F+D  & No F+D          &&        F+D  & No F+D  &&        F+D  & No F+D &&      F+D  &  No F+D && \\
    \hline
    \textsc{Addition}        &&  \\
    \hline
    normalized      &&     0.53  &      0.53   &&    0.53  &     0.53   &&    0.39  &     0.39  &&    0.87  &     0.87  &&  C  \\
    normalized      &&     0.25  &      0.25   &&    0.25  &     0.25   &&    0.25  &     0.25  &&    0.26  &     0.26  &&  ASM  \\
    overflow        &&     0.25  &      0.25   &&    0.25  &     0.25   &&    0.25  &     0.25  &&    0.26  &     0.26  &&  ASM  \\
    underflow       &&     0.25  &     38.20   &&    0.25  &    37.70   &&    0.25  &    31.90  &&    0.26  &    36.30  &&  ASM  \\
    denormal l      &&     0.25  &      0.25   &&    0.25  &     0.25   &&    0.25  &     0.25  &&    0.26  &     0.26  &&  ASM  \\
    denormal r      &&     0.25  &      0.25   &&    0.25  &     0.25   &&    0.25  &     0.25  &&    0.26  &     0.26  &&  ASM  \\
    both denormals  &&     0.25  &      0.25   &&    0.25  &     0.25   &&    0.25  &     0.25  &&    0.26  &     0.26  &&  ASM  \\
    NaN             &&     0.25  &      0.25   &&    0.25  &     0.25   &&    0.25  &     0.25  &&    0.26  &     0.26  &&  ASM  \\
    \hline
    \textsc{Multiplication} &&    \\
    \hline
    normalized      &&     0.56  &      0.56   &&    0.54  &     0.54   &&    0.39  &     0.39  &&    0.88  &     0.88  &&   C  \\
    normalized      &&     0.25  &      0.25   &&    0.25  &     0.25   &&    0.13  &     0.13  &&    0.26  &     0.26  &&  ASM  \\
    overflow        &&     0.25  &      0.25   &&    0.25  &     0.25   &&    0.13  &     0.13  &&    0.26  &     0.26  &&  ASM  \\
    underflow       &&     0.25  &     40.00   &&    0.25  &    39.50   &&    0.13  &    32.70  &&    0.26  &    36.50  &&  ASM  \\
    denormal l      &&     0.25  &     36.20   &&    0.25  &    35.70   &&    0.13  &    32.70  &&    0.26  &    37.60  &&  ASM  \\
    denormal r      &&     0.25  &     36.20   &&    0.25  &    35.70   &&    0.13  &    32.70  &&    0.26  &    37.60  &&  ASM  \\
    both denormals  &&     0.25  &      0.25   &&    0.25  &     0.25   &&    0.13  &     0.13  &&    0.26  &    37.60  &&  ASM  \\
    NaN             &&     0.25  &      0.25   &&    0.25  &     0.25   &&    0.13  &     0.13  &&    0.26  &     0.26  &&  ASM  \\
    \hline
    \textsc{Division} &&   \\
    \hline
    normalized      &&    11.10  &     11.10   &&    7.10  &     7.10   &&    7.08  &     7.08  &&    4.97  &     4.97  &&    C  \\
    normalized      &&    10.30  &     10.30   &&    6.68  &     6.68   &&    6.65  &     6.65  &&    4.90  &     4.90  &&  ASM  \\
    overflow        &&    11.00  &     11.00   &&    7.01  &     7.01   &&    7.02  &     7.02  &&    4.90  &     4.90  &&  ASM  \\
    underflow       &&    11.00  &     71.10   &&    7.01  &    63.30   &&    7.02  &    56.50  &&    4.90  &    41.40  &&  ASM  \\
    div-by-zero     &&     5.04  &      5.04   &&    4.13  &     4.13   &&    4.04  &     4.04  &&    2.02  &     2.02  &&  ASM  \\
    denormal dividend
                    &&     5.04  &     64.30   &&    4.13  &    57.00   &&    4.04  &    54.00  &&    2.02  &    42.50  &&  ASM  \\
    denormal divisor&&     5.04  &     64.30   &&    4.13  &    57.00   &&    4.04  &    54.00  &&    2.02  &     4.90  &&  ASM  \\
    both denormals  &&     5.04  &     64.30   &&    4.13  &    57.00   &&    4.04  &    54.00  &&    2.02  &     4.90  &&  ASM  \\
    \hline
    \textsc{Fused-Multiply-Add}       &&  \\
    \hline
    normalized      &&           &             &&          &            &&    0.22  &     0.22  &&    0.26  &     0.26  &&  ASM  \\
    multiplication overflow
                    &&           &             &&          &            &&    0.22  &     0.22  &&    0.26  &     0.26  &&  ASM  \\
    addition overflow
                    &&           &             &&          &            &&    0.22  &     0.22  &&    0.26  &     0.26  &&  ASM  \\
    multiplication underflow
                    &&           &             &&          &            &&    0.22  &     0.22  &&    0.26  &     0.26  &&  ASM  \\
    addition underflow
                    &&           &             &&          &            &&    0.22  &    33.50  &&    0.26  &    36.50  &&  ASM  \\
    \hline
  \end{tabular}
  \caption{
    Duration of a single floating-point operation in cycles for specific AVX
    floating point benchmarks with different values of the input operands.
    The visible duration of the full AVX instruction is four times the reported
    number, which is also the inverse throughput.
    Measurements were obtained with FTZ and DAZ enabled and disabled, denoted as
    F+D and No F+D, respectively.
    In the assembly benchmarks (ASM) the vectors were kept in registers to avoid
    the load/store bottleneck.
  }
  \label{tab:results}
  \end{center}
\end{sidewaystable}

\section{Conclusion}

Floating point operations like addition and multiplication with normalized input 
and output values are handled in three to five cycles.
With enabled flush-to-zero (FTZ) and denormals-are-zero (DAZ), which is the
default case for GCC and the Intel compiler if not otherwise specified,
underflow, overflow, NaNs, and divisions-by-zero have no negative performance
impact.

If however, the additional precision gained by gradual underflow is required 
FTZ and DAZ must be disabled.
The costs for underflowing operations are then about two magnitudes higher than 
the normalized operations for AVX addition, multiplication, and division.
In the case of FMA only an underflow during the addition is costly.
An underflowing multiplication within FMA introduces no additional costs. 

\bibliographystyle{abbrv}
\bibliography{Bibliography}

\begin{thebibliography}{1}

\bibitem{ieee-754-2008}
{IEEE} standard for floating-point arithmetic.
\newblock {\em IEEE Std 754-2008}, pages 1--70, Aug 2008.
\newblock \bibdoiurl{10.1109/IEEESTD.2008.4610935}.

\bibitem{amd-guide-2014}
{AMD Inc.}
\newblock {S}oftware {O}ptimization {G}uide for {AMD} {F}amily 15h
  {P}rocessors.
\newblock \url{http://support.amd.com/TechDocs/47414_15h_sw_opt_guide.pdf},
  2014.

\bibitem{fog-2014}
A.~Fog.
\newblock Instruction tables: {L}ists of instruction latencies, throughputs and
  micro-operation breakdowns for {I}ntel, {AMD} and {VIA} {CPU}s.
\newblock \url{http://www.agner.org/optimize/instruction_tables.pdf}, 2014.

\bibitem{intel-ftz-daz}
{Intel Corp.}
\newblock x87 and {SSE} floating point assists in {IA}-32: Flush-to-zero
  ({FTZ}) and denormals-are-zero ({DAZ}).
\newblock
  \url{https://software.intel.com/en-us/articles/x87-and-sse-floating-point-assists-in-ia-32-flush-to-zero-ftz-and-denormals-are-zero-daz},
  2008.

\bibitem{intel-orm-2014}
{Intel Corp.}
\newblock {I}ntel64 and {IA}-32 {A}rchitectures {O}ptimization {R}eference
  {M}anual.
\newblock
  \url{http://www.intel.com/content/dam/doc/manual/64-ia-32-architectures-optimization-manual.pdf},
  2014.
\newblock Version: April 2014.

\end{thebibliography}

\end{document}